\documentclass[aip,pre,twocolumn,superscriptaddress,reprint]{revtex4-2}

\usepackage{amsfonts}
\usepackage{amsmath}
\usepackage{amssymb}
\usepackage{graphicx}
\usepackage{float}
\usepackage[dvipsnames]{xcolor}
\begin{document}

\title{Activation force dependence of the  fluidization transition in supercooled active water%:a decoupling between temperature and cooperativity.
}

\date{{\color{black}\today} }

\author{Quoc Tuan Truong}
\affiliation{Laboratoire de Photonique d'Angers EA 4464, Universit\' e d'Angers, Physics Department,  2 Bd Lavoisier, 49045 Angers, France.}

\author{Victor Teboul}
\email{victor.teboul@univ-angers.fr}
\affiliation{Laboratoire de Photonique d'Angers EA 4464, Universit\' e d'Angers, Physics Department,  2 Bd Lavoisier, 49045 Angers, France.}


\begin{abstract}
Using an activation model reproducing facilitation mechanisms, we found recently a fluidization transition in supercooled water triggered by the motion of a few active molecules. We investigate here the effect of the activation force on the transition.
The activation force increases the probability for active molecules to overcome the energy barrier induced by caging. 
The induced molecular motions result in an intermittent loose structure leading to a decrease of energy barriers around the moving molecule, thus facilitating the motion of nearby molecules.
Therefore we expect the force, if small enough, to increase the facilitation mechanism, and as a result cooperative motions induced by it.
As a result the activation force permits to tune facilitation and cooperative motions in active glass-formers.
 We find an optimum force for cooperative motions that we evaluate to be around $3.5 pN$, a value in the range of protein motor forces. 
As expected for a phase transition we observe peaks of cooperative motions (dynamic heterogeneity) on the transition. This result shows that dynamical heterogeneity are the cooperative motions of physical importance for the transition. %The same conclusion applies probably also to the glass transition mechanism due to the similarity between the two transitions.

%Results:\\
%1. Optimum force for cooperativity.\\
%2. Peaks of cooperativity on the transition.\\
%3. Power law relation between concentration and force with a change in slope.\\
%4. Effect of various cooperativity values on the diffusion and alpha relaxation time.\\
%. Possible relation with glass-transition.\\

\end{abstract}

\maketitle
{
\section{ Introduction}
Media containing self-propelled particles, such as molecular motors, motor proteins or synthetic motors, are called active\cite{active1,active2,active3,active4,active5,active6,active7,active8,active9,active10,mandal,motoro1,motoro2,motoro3,motoro4,motoro5,motoro6,Szamel1,Szamel2,Szamel3,Szamel4,Szamel5,Szamel6,Szamel7,meta}.
Using an activation model that reproduces facilitation mechanisms\cite{facile,facile1,facile2,facile3,facile5,facile6}, we found in previous studies, activation induced phase transitions\cite{a1,a2,a12} between viscous supercooled liquids and fluidized versions of these liquids that nonetheless retain cooperative motion\cite{dh0,dh1,dh2,dh3,iso1,iso2,silica,prefold}. 
The fluidization transition was first identified in a model supercooled liquid\cite{ariane,a1,a2}, and later observed in supercooled water\cite{a12} where it triggers in some conditions a liquid-liquid structural transition\cite{LDL} between low density liquid water (LDL) and high density liquid water (HDL).
Studies specializing in water\cite{wat1,wat3,usb1,wat0b,wat2,wat0a,wat0} and active water\cite{a12} are of particular interest due to the importance of water in biology, to the quite specific properties of water and to the liquid-liquid transition\cite{LDL,LDLS1,LDLS2,LDLS3,LDLS4,LDLS5,LDLS6,LDLS7,LDLS8,LDLS9,LDLS10,LDLS11,LDLA3,LDLE1,LDLE2,LDLE3,LDLE4,LDLE5,LDLE6,LDLE7,LDLE8,LDLA1,LDLA2,LDLN,LDLN2}.
In this work we investigate the effect of a variation of the activation force on the fluidization transition.
The activation force is an important parameter that has two main effects. The first effect of the force is to increase the kinetic energy of the molecule while it travels inside the cage constituted of its neighbors, therefore helping cage breaking and diffusion. 
The force increase can, in this perspective, be seen as equivalent to a local and temporary temperature increase. The second effect is an increase of the facilitation mechanism, due to the rearrangement around the moving active molecule that facilitates the motion of neighbors. This leads to an increase of cooperative motions, i.e. dynamical heterogeneity (DHs). 
Both effects contribute to the fluidization transition as they both ultimately increase the diffusion coefficient and cooperative motions.\\

This is in contrast with the glass transition for which diffusion (temperature) and cooperative motions (dynamic heterogeneity)  vary in opposition, as spontaneous thermal DHs increase, together with the facilitation when the temperature decreases.
There are however a lot of similarities between the fluidization transition observed here and the glass transition, as both can be seen as fluidization transitions controlled by the concentration of excitations, but the facilitation mechanism is acting in opposite ways in the two transitions.

The activation force parameter therefore may ultimately permit to separate the facilitation mechanism effect from the temperature effect on the material.

%The effect of the force could then be seen as decreasing the potential energy barrier or more precisely increasing locally the temperature.
 %In this picture we may model the force effect with an effective temperature. However there are two main differences between the temperature effect and the force effect. The first difference is that the force doesn't fluctuates in our model, while the temperature corresponds to a distribution of kinetic energies. The second difference is that the force is expected to increase the facilitation mechanism, while the temperature mostly decreases it.
%Therefore we expect an increase of cooperative mechanisms with the activation force in contrast with the temperature effect.
%Modifying the force therefore permits us to discriminate dynamical properties due to cooperative motions from the properties due to the temperature, that is the concentration of excitations $c_{ex}=c_{0} exp(-E_{a}/kT)$ where $Ea$ is the activation energy needed to overthrow the potential wall induced by the cage effect.

%As the activation force increases we observe successively two different dynamical regimes. 
%For small forces, the medium is only slightly perturbed and cooperative motions increase rapidly with the force.
%Above a force threshold the medium begins to be more strongly perturbed leading to a modification of the dynamics.

\section{Calculation}

Due to water's ubiquity on earth, its importance to biology and its complex behavior, a number of intermolecular potentials have been developed to model water\cite{wpot1,wpot2,wpot3,wpot4,wpot5,wpot6,wpot7,wpot8,wpot8b,wpot8c,wpot8d}. In this work, we model the molecular interactions of water using the TIP5PE potential\cite{wpot4,wpot5,wpot6}, and we treat long-range electrostatic interactions with the reaction field method\cite{md1,md2,md3}, employing a cutoff radius $R_{RF}=9$ Å and an infinite dielectric constant for distances $r>R_{RF}$. The water molecule is modeled as a rigid body, and our analysis focuses on the behavior of the molecular center of mass (which closely approximates the oxygen position, since the difference between the two is small in water).
The equations of motion are integrated using the Gear 4 algorithm\cite{md1,md2,md3} with a time step $\Delta t = 10^{-15}$ s.\\

Active matter has attracted considerable interest\cite{active1,active2,active3,active4,active5,active6,active7,active8,active9,active10,mandal,motoro1,motoro2,motoro3,motoro4,motoro5,Szamel1,Szamel2,Szamel3,Szamel4,Szamel5,Szamel6,Szamel7,meta} due to its relevance to biology and nonequilibrium statistical physics. Active matter also provides a novel framework for addressing the long-standing glass-transition problem\cite{gt0,gt1,gt2,anderson,fragile1}, owing to its nonequilibrium nature and its potential underlying mechanisms. %Studies of supercooled active matter have shown that introducing active particles into a viscous medium reduces its viscosity without significantly increasing the temperature, an effect known as fluidization.
Notice that experimental technics like interaction induced light scattering\cite{usb2,usb3,usb4} and interaction induced absorption\cite{usb4,usb5}, usually employed for the improvement of interaction potentials and polarizabilities, could probably lead to better understanding of active matter and cooperative motions. That line of research has however never been opened.

Because, in supercooled liquids, the cage effect randomizes velocity directions, velocity is not a relevant parameter at low temperature. We therefore use the mobility to define the active force instead. We define the mobility as
%\begin{equation}
%\mathbf{\mu_i}(t,\tau_m)=\mathbf{r_i}(t+\tau_m)-\mathbf{r_i}(t),
%\label{e0}
%\end{equation}

%where $\tau_m$ is a characteristic time associated with cooperative motion.

%A  thermostat\cite{berendsen}  removes the energy dissipated into the system avoiding a  drift in energy. 
%Due to cage effect that randomizes the velocity direction, velocity is not a relevant parameter in supercooled liquids. Therefore, we use the mobility instead to define the active force. We define the mobility  as:

\begin{equation}
\displaystyle{{\bf \mu_{i}}(t,\tau_{m})= {{\bf r_{i}} (t+\tau_{m})-{\bf r_{i}}(t)}    }       \label{e0}
\end{equation}
Where $\tau_{m}$ is a characteristic time that defines the mobility and is related to cooperative motions dynamics.

 In our simulations, a concentration $C$ of the molecules (typically a few percent) is activated: during a time interval of $T_a = 10$ ps, the activated molecules are pushed in the mobility direction of the most mobile among their neighbors; activation is then switched off for 30 ps, giving a total periodicity of $T_p = 40$ ps. On average, a fraction $C/4$ of the molecules is activated at any given time, which corresponds to 2.5 \% at the transition, while the remaining molecules experience only intermolecular forces.
$T_a$ serves as the persistence time in our study, so increasing the ratio $T_a/T_p$ increases the level of activation, while $\tau_m$ characterizes cooperative mechanisms.

%In our calculations,  a concentration $C$ of the medium molecules (typically a few percents) are activated (pushed) in the mobility direction of the most mobile of their neighbors during $T_{a}=10 ps$ then the activation stops during $30ps$ leading to a  $T_{p}=40 ps$ periodicity. Therefore, an average of $C/4$ molecules are activated,  that is $2.5 \%$ at the transition, while the other molecules do not experience external forces but only the intermolecular interactions. 
 %$T_{a}$ is the persistence time in our study, and an increase of the ratio $T_{a}/T_{p}$ will lead to an increase of the activation, while $\tau_{m}$ relates to the cooperative mechanisms.

 A thermostat\cite{berendsen} removes the energy dissipated into the system, preventing any drift in total energy.
 
 We investigate how key features of supercooled liquids change under activation, including dynamic heterogeneity, diffusion, and the $\alpha$-relaxation time, which is associated with the viscosity of the medium. We now define the statistical functions used in this analysis. A central quantity in glass-transition studies is the intermediate scattering function, $F_S(Q,t)$, which describes the autocorrelation of density fluctuations at wave vector $Q$. It probes the structural relaxation of the material and is defined as

\begin{equation}
\displaystyle{F_{S}(Q,t)={1\over N N_{t_{0}}} Re( \sum_{i,t_{0}} e^{i{\bf Q.(r_{i}(t+t_{0})-r_{i}(t_{0}))}}  )          }\label{e1}
\end{equation}
For physical reasons, Q is chosen as the wave vector (here $Q_{0}=2.25$\AA$^{-1}$) corresponding to the maximum of the structure factor $S(Q)$.
$F_{S}(Q_{0},t)$ then allows us to calculate the $\alpha$ relaxation time $\tau_{\alpha}$  of the medium from the equation: 
\begin{equation}
\displaystyle{F_{S}(Q_{0},\tau_{\alpha})=e^{-1}}       \label{e10}     
\end{equation}
Finally, the diffusion coefficient $D$ is obtained from the long time limit of the mean square displacement $<r^{2}(t)>$:
\begin{equation}
\displaystyle{<r^{2}(t)>=   {1\over N N_{t_{0}}}  \sum_{i,t_{0}}  ({\bf r}_{i}(t+t_{0})-{\bf r}_{i}(t_{0}))^{2}                                }       \label{e11}     
\end{equation}
and
\begin{equation}
\displaystyle{\lim_{t \to \infty}  <r^{2}(t)>=  2 d D t                               }       \label{e12}     
\end{equation}

In this work we will chose $\tau_{m}=4 ps$ unless otherwise specified and the dimension $d=3$.

%\vskip0.4cm

\section{Results and discussion}

In a previous work we found a  fluidization transition triggered by active molecules in a supercooled model liquid\cite{a1,a2,ariane}, we then observed this transition in supercooled water\cite{a12} using the concentration of active molecules as control parameter. %We understand this transition as arising from the aggregation of active molecules induced by facilitation, that then transfer their mobility to non-active ones.

In this work we study the effect of the activation force on the transition.
A temporary reorganization of molecules appears near moving molecules due to cage breaking.
Due to this temporary reorganization, motion is facilitated around moving molecules, a process called facilitation. 
Facilitation then induces an aggregation of moving molecules moving cooperatively.
Because active molecules increase cage breaking mechanisms around them when their activation force increases, we expect the force to increase facilitation mechanisms for non active molecules and as a result cooperative motions.
We will therefore now investigate the effect of the force on cooperative motions. %For that purpose we will use different tools to measure cooperative motions, beginning with the Non Gaussian parameter (NGP).

\subsection{Force-Cooperativity}

\subsubsection{Peak on the transition}

To evaluate the evolution of cooperative motions, figure \ref{fig1x} shows the evolution of the maximum value of the Non Gaussian parameter $\alpha_{2}(t^{*})$ as the force varies, while the concentrations of active molecules is kept constant. The different figures correspond to different values of the concentration $C$. 
For small forces, the NGP remains roughly equal to its thermal (non activated) value $\alpha_0$, then when we reach the phase transition we observe a peak of cooperativity, followed at larger forces by a decrease to values that remain larger than  $\alpha_0$.
The peak of cooperativity on the critical point is expected for a phase transition\cite{book1,book2,book3}. It shows that cooperativity is necessary for the transition to occur, in order to transmit motion from active molecules to the whole medium.
%As the concentration of active molecules or the force increases, the number of moving molecules increases until ... ??? 
Below the transition, the number of molecules moving cooperatively appears approximately not affected by the activation, while above the transition the NGP is significantly larger than without activation.

\begin{figure}%[H]
\centering
\includegraphics[height = 5.8 cm]{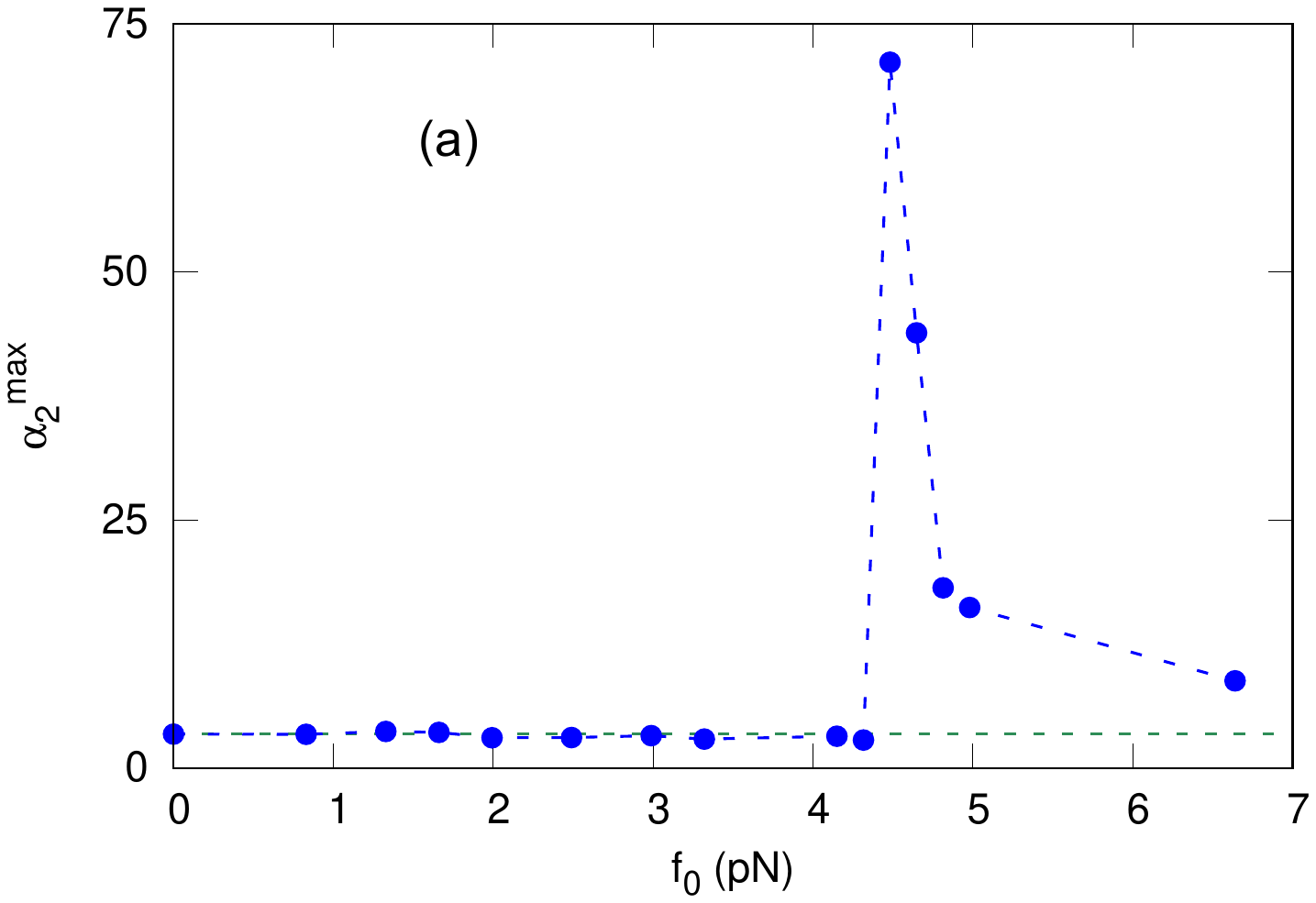}
\includegraphics[height = 5.8 cm]{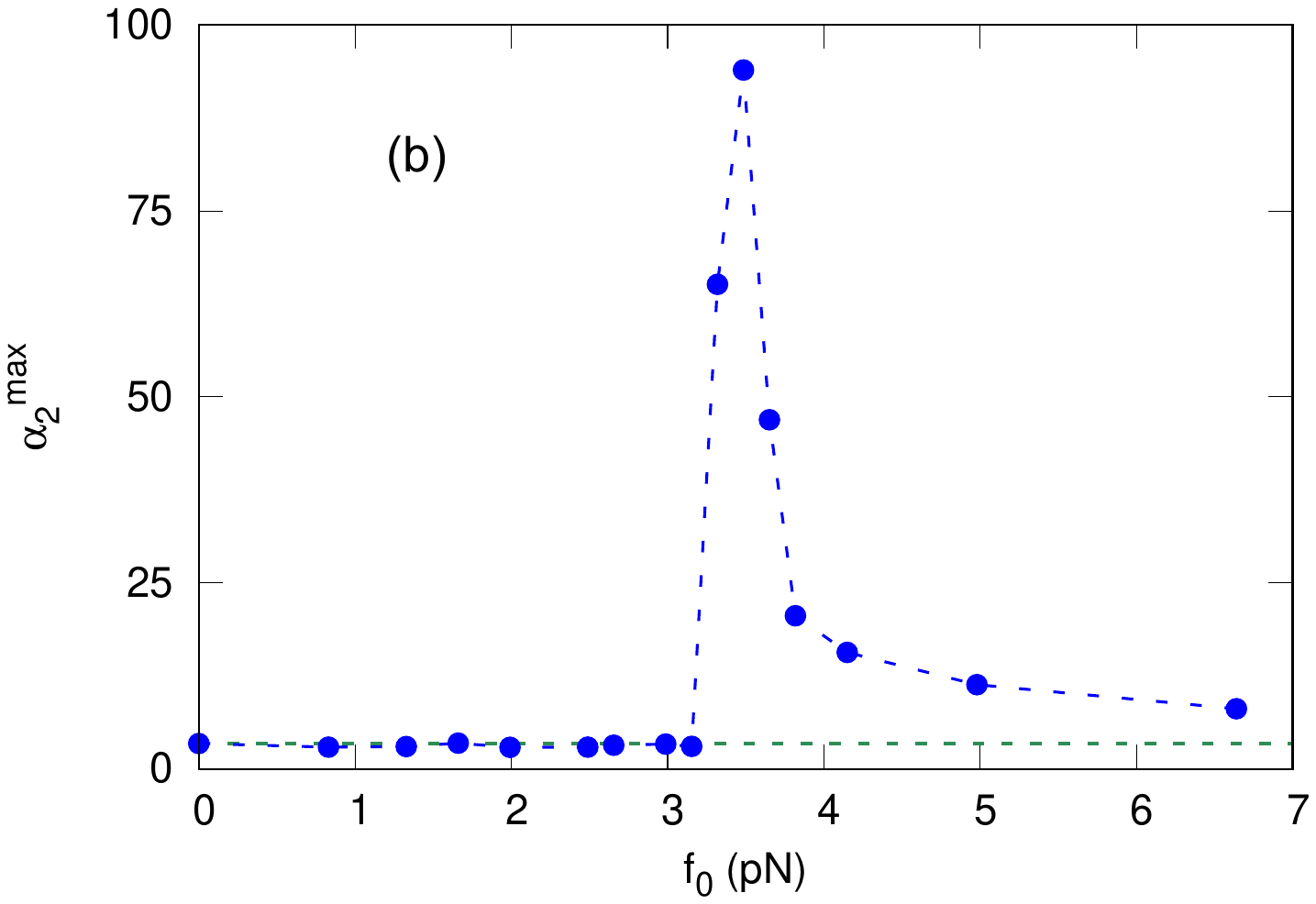}
\includegraphics[height = 5.8 cm]{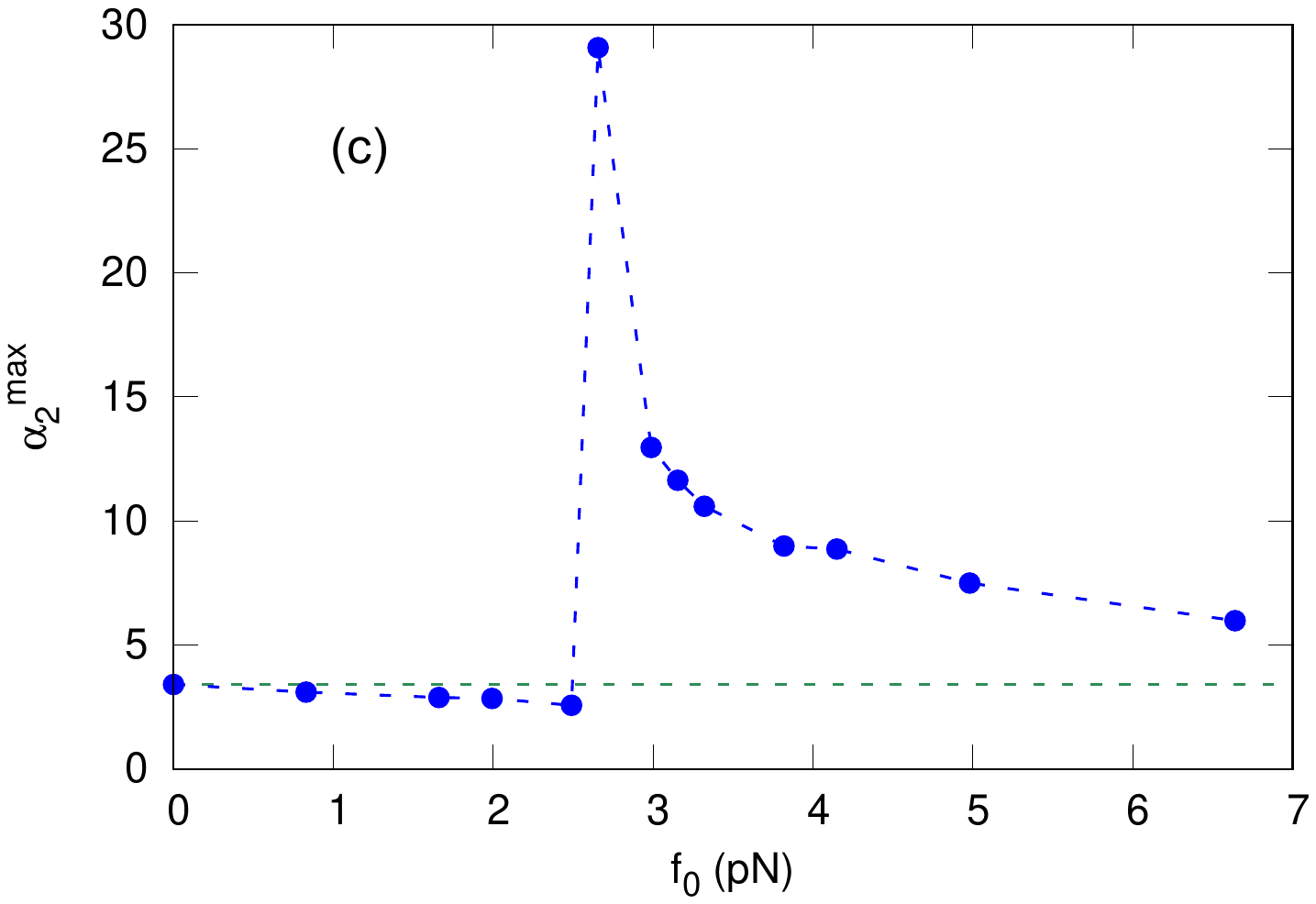}

\caption{(color online) Maximum of the Non Gaussian parameter $\alpha_{2}(t)$ versus the activation force $f_0$, for a constant concentration of active molecules (a) $C=5\%$, (b) $C=10\%$, (c) $C=20\%$. T=220 K. The peak corresponds to the transition. Below the transition (left part of the figure) the cooperativity decreases slightly from the spontaneous one observed without activation ($f_{0} = 0 pN$). Above the transition (right part of the figure) it decreases from the peak at the transition. Other concentrations lead to similar plots, with the peak shifted.
}
\label{fig1x}
\end{figure}

We show in Figure \ref{fig4x} the coefficient $K=D.\tau_{\alpha}$ measuring the breaking of the Stokes-Einstein law, as a function of the concentration of active molecules for various forces.
The breaking of the Stokes-Einstein law is due to the rise of cooperative motions called dynamic heterogeneities in supercooled liquids and can therefore be used to measure cooperative motions. The relation between dynamic heterogeneities, the Non Gaussian parameter and the breaking of Stokes-Einstein's law in supercooled liquids has been investigated in a number of works.
We observe peaks of cooperativity (measured by $K$) on the fluidization transitions. 
Below the transition, the cooperativity is small, then at the transition there is a peak followed by an increase of cooperative motions above the transition.
The peak indicates a direct relation between the rise of cooperativity and the transition.
Phase transitions are usually related to a sharp increase in cooperative mechanisms,  permitting the transition to propagate from a microscopic fluctuation to a macroscopic transition. However our results (displayed in Figures \ref{fig1x} and \ref{fig4x}) show that the measured cooperative motions (dynamic heterogeneities) are the ones physically significant for the transition. 

The increase of cooperative motion above the transition suggests that the fluidization needs cooperative motions to persist as the temperature is low.
Cooperative motions are larger for larger concentration of active molecules as active molecules generate them.

\begin{figure}%[H]
\centering
\includegraphics[height = 7 cm]{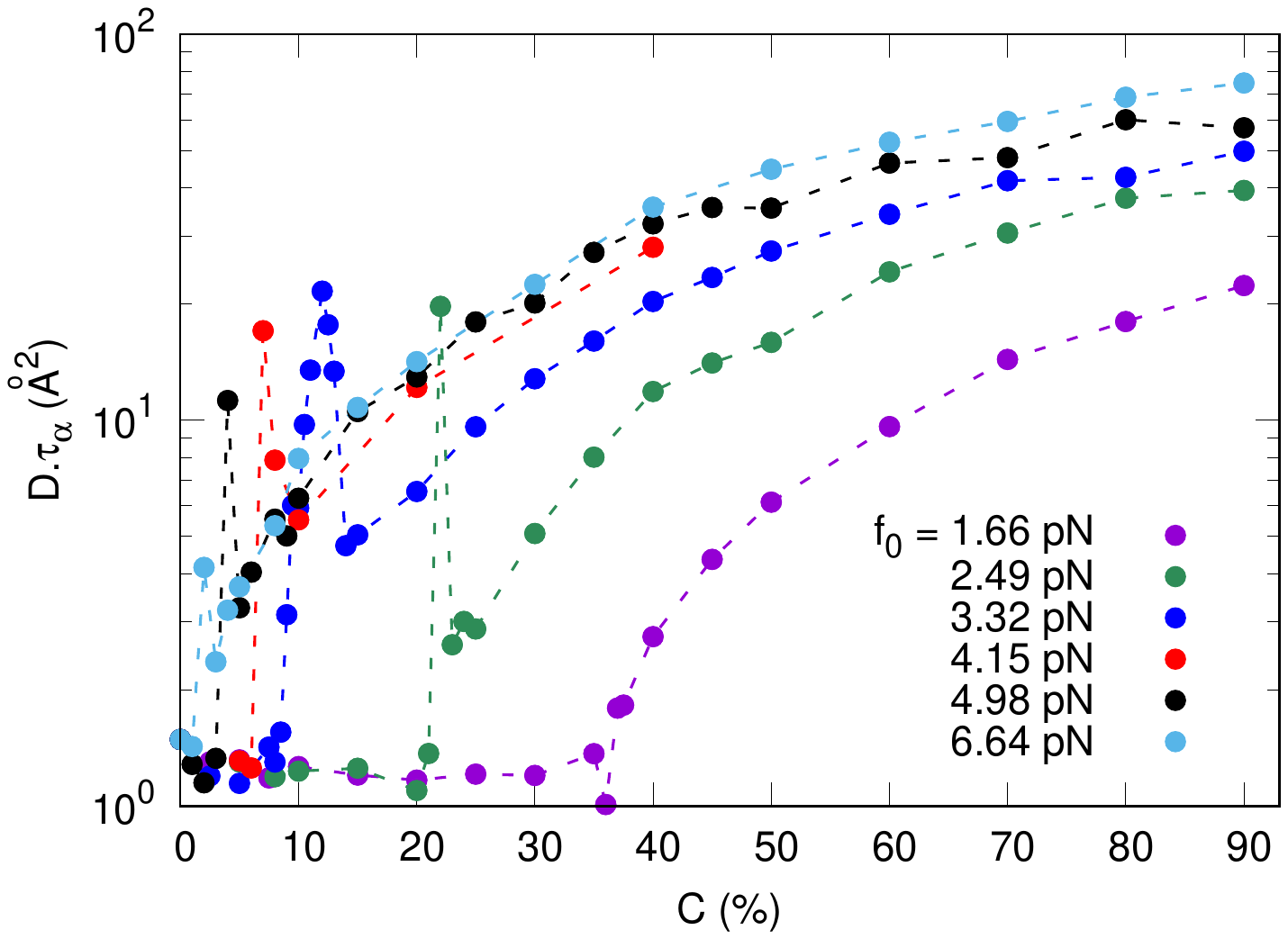}
	
\caption{(color online) Stokes Einstein breaking, measured with $K=D.\tau_{\alpha}$ coefficient $K$ at constant temperature $T=220 K$.
Stokes Einstein Law predicts a constant coefficient $K$ and deviation from the law are associated to the rise of cooperativity in supercooled liquids.
The coefficient at small concentration that is of the order of unity, rises at the transition. A peak of cooperativity is observed at the transition, as expected for a phase transition. The peaks (theoretically infinites) are here limited in height due to the limited size of the system. The larger peak appears for $f_0=3.32 pN$. 
}
\label{fig4x}
\end{figure}

The rise of cooperative motions can also be seen directly on the Self Van Hove correlation function $G_{s}(r,\Delta t)$ in Figure \ref{fig31}.
$G_{s}(r,\Delta t)$ represents the density probability to find a molecule a distance r apart its previous position after a time lapse $\Delta t$.
Cooperative motions lead to a tail in the Van Hove for large values of r, as cooperatively moving molecules move to larger distances than regular ones.
Figure \ref{fig31} shows an increase of the tail as the force is increased, showing the increase of cooperative motions with the activation force.

\begin{figure}%[H]
\centering
\includegraphics[height = 7 cm]{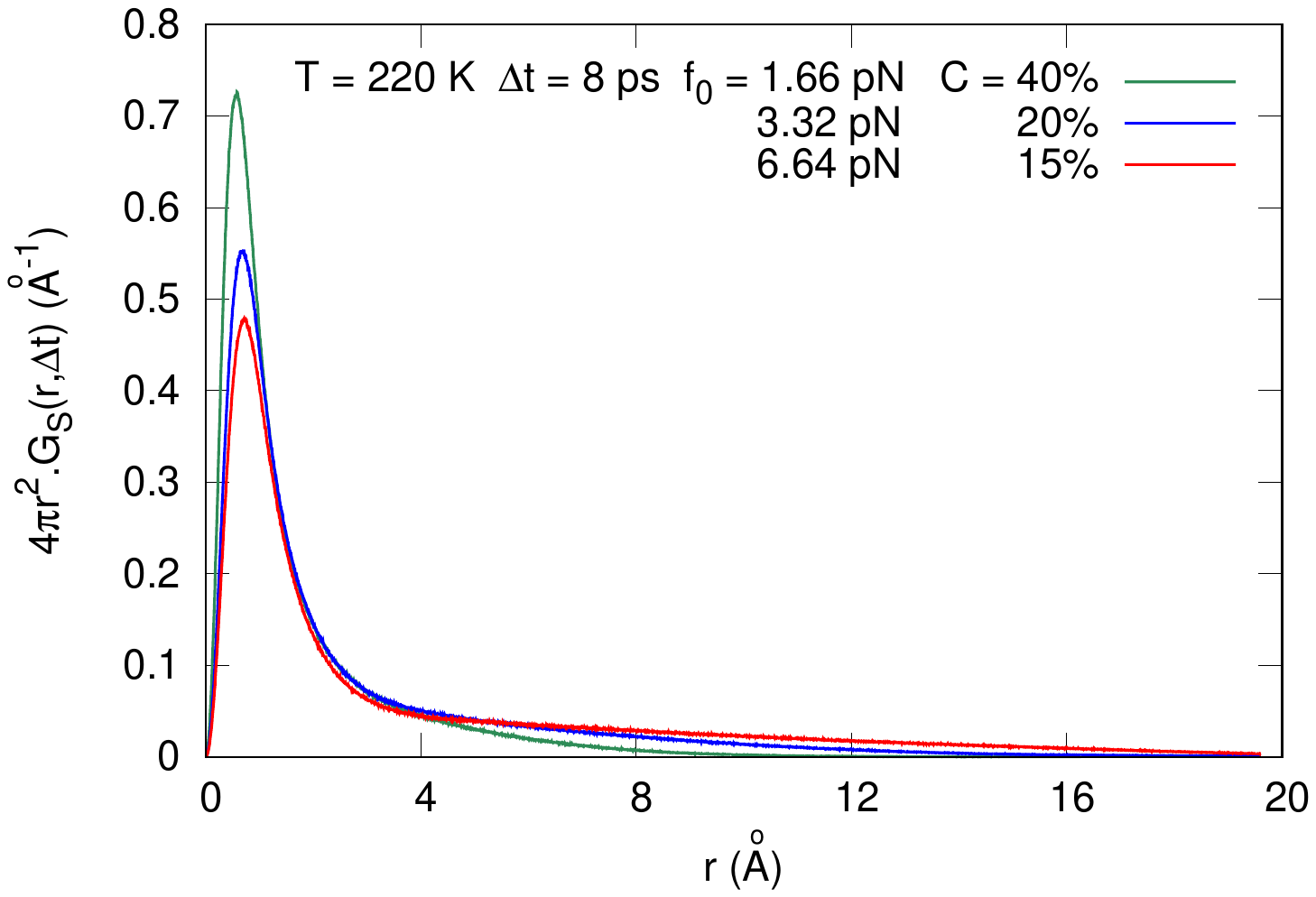}
	
\caption{(color online) Van Hove self correlation function $G_{s}(r,\Delta t)$ with $\Delta t = 8 ps$ displayed for various forces $f$. 
Concentrations are chosen to be above the transition with the corresponding forces. We observe the increase of the tail characteristic of cooperative motions called dynamic heterogeneities, when the force increases.
}
\label{fig31}
\end{figure}

\subsubsection{Optimum force value for facilitation}

Is facilitation always increasing with the activation force, or is there an optimum force for facilitation, and therefore for the fluidization transition ? To answer that question we plot in Figure \ref{fig2x} the maximum value of  the non Gaussian parameter $\alpha_{2}(t^{*})$ on the transition versus the activation force. 
The curve shows an optimum force around $3.2 pN$ for facilitation or cooperative motions. 
We observe the same behavior in Figure \ref{fig3x} with the Stokes Einstein breaking, however with a wider maximum centered around $3.5 pN$.
We interpret that behavior as follows. When the force increases, the facilitation increases due to an increased structural perturbation around the moving molecule.
The structural perturbation facilitates the motion of nearby molecules due to loose caging.
This trend continues as long as the force doesn't modify permanently the medium. When the force affecting the medium permanently is reached, the facilitation begins to decrease. As a result we observe an optimum force and a maximum value for facilitation. 
We observe in Figures  \ref{fig2x} and \ref{fig3x} that at the optimum force, the cooperative motions are quite enhanced.
Notice that the optimum force $\approx 3.5 pN$ is within the range of forces found in biological molecular motors.

\begin{figure}%[H]
\centering
\includegraphics[height = 7 cm]{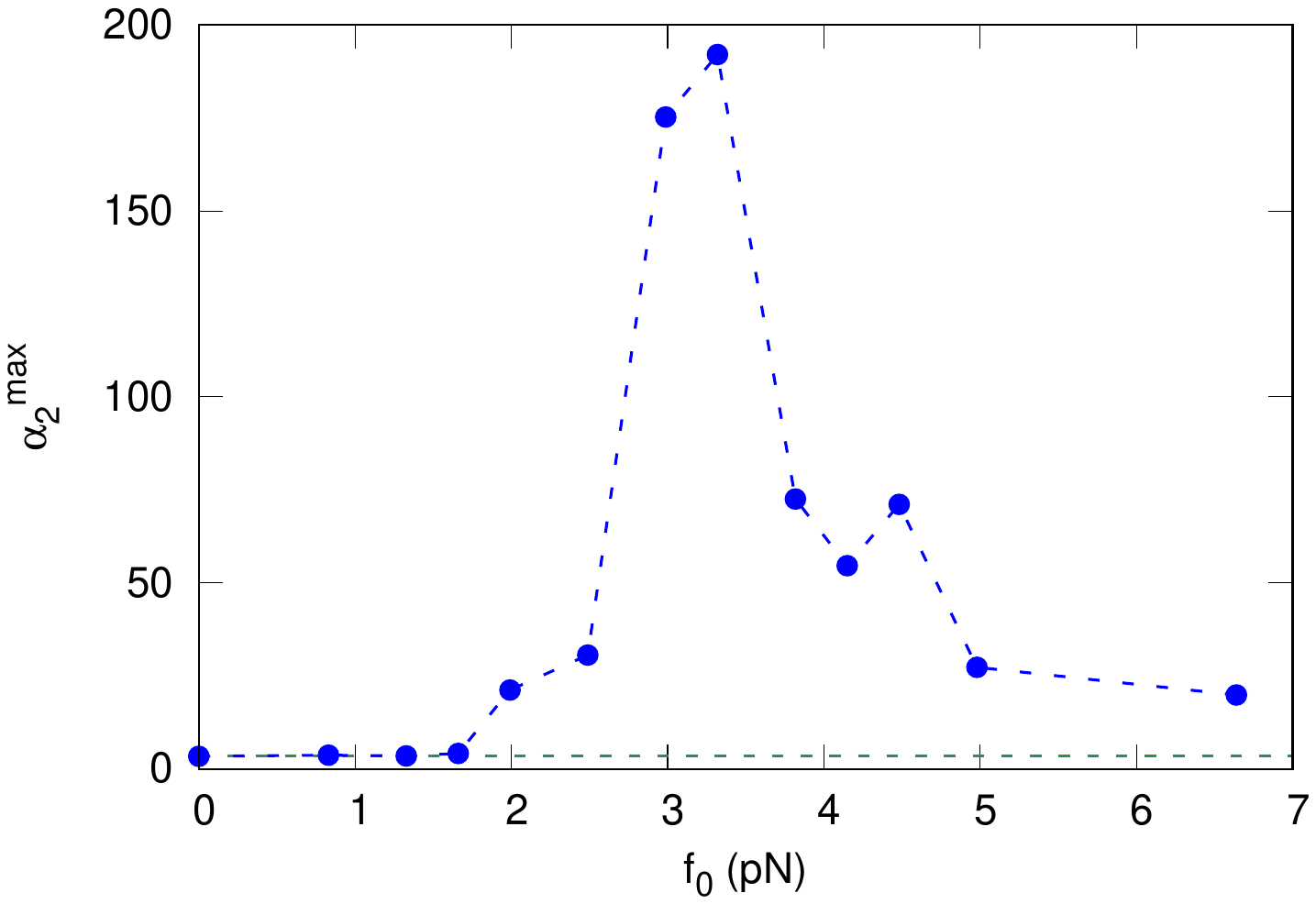}
\caption{(color online) Maximum of the Non Gaussian parameter $\alpha_{2}(t)$ on the transition critical point versus the activation force $f_0$. T=220 K.
The concentration of active molecules is chosen here for each force to be on the critical point. We found that the position of the peak doesn't vary with the temperature, while the height of the peak decreases when the temperature increases.
}
\label{fig2x}
\end{figure}

\begin{figure}%[H]
\centering
\includegraphics[height = 7 cm]{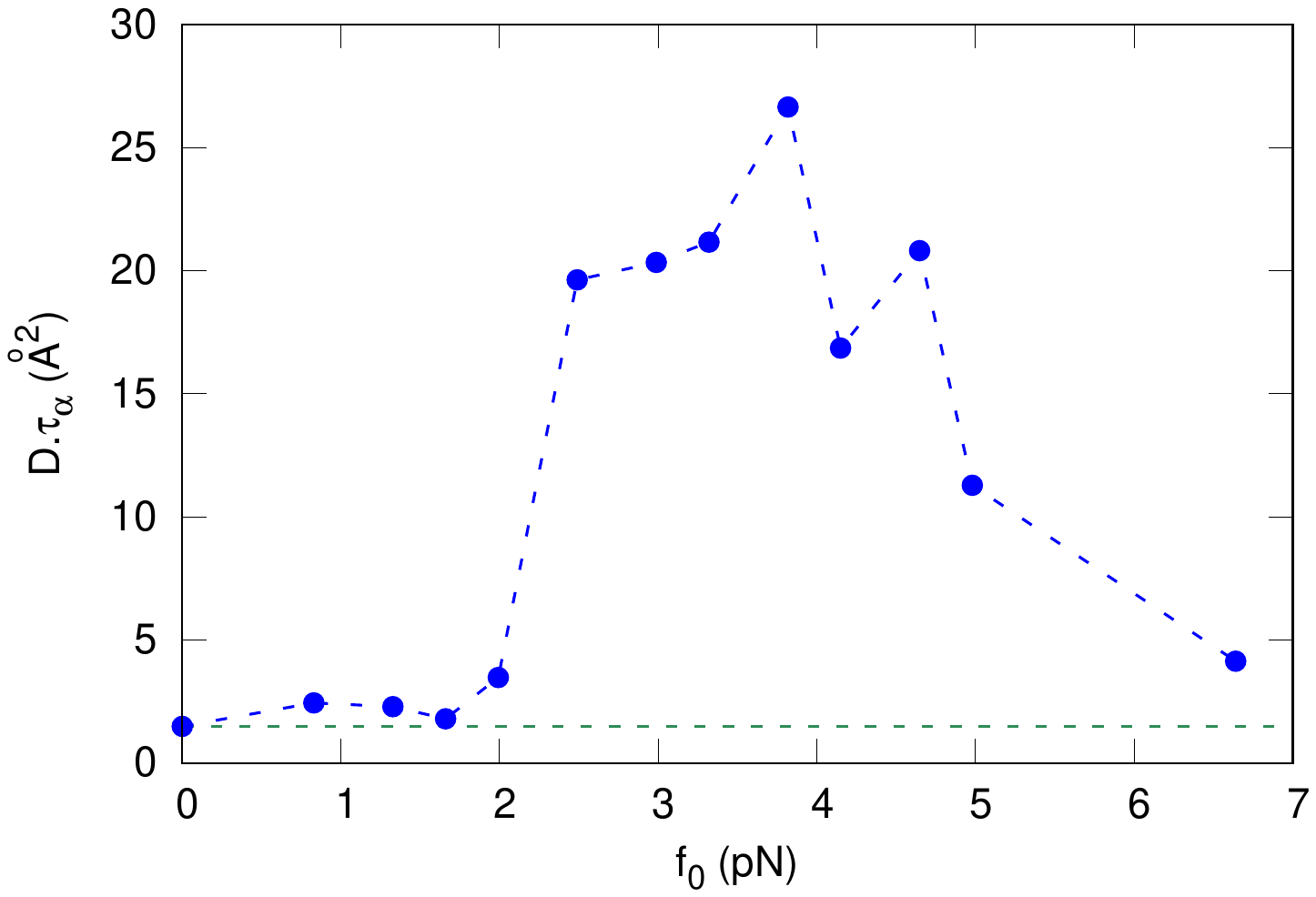}
	
\caption{(color online) Maximum of $D.\tau_{\alpha}$ coefficient on the transition critical point versus the activation force $f_0$. T=220K. Concentrations are critical concentrations corresponding to each force displayed. }
\label{fig3x}
\end{figure}

%\subsection{Force-concentration}
\subsubsection{Cooperative motions effect on the transition}

\begin{figure}%[H]
\centering
\includegraphics[height = 7 cm]{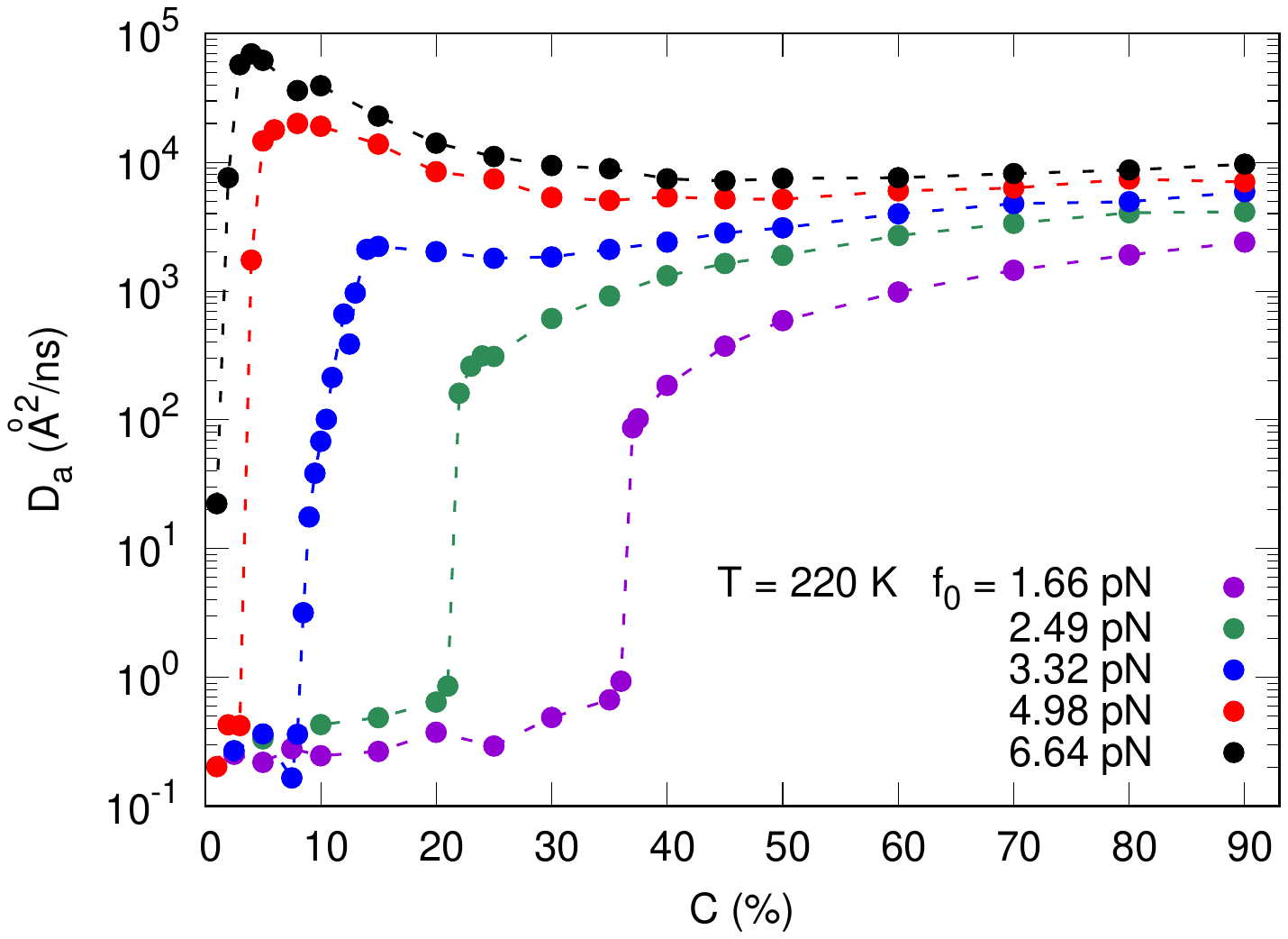}
\caption{(color online) Diffusion coefficient D of active molecules as a function of the concentration $C$ of active molecules for various activation  forces. For forces equal to the optimum or larger a bump develops around the transition, probably induced by cooperative mechanisms as it is not present for the alpha relaxation time. %The dashed line is a guide to the eye that corresponds to $C = 24\%$. %Below the fluidization transition, the diffusion coefficient strongly depends on temperature, whereas above the transition, the data nearly collapse onto a single curve.
}
\label{fig01a}
\end{figure}

\begin{figure}%[H]
\centering
\includegraphics[height = 7 cm]{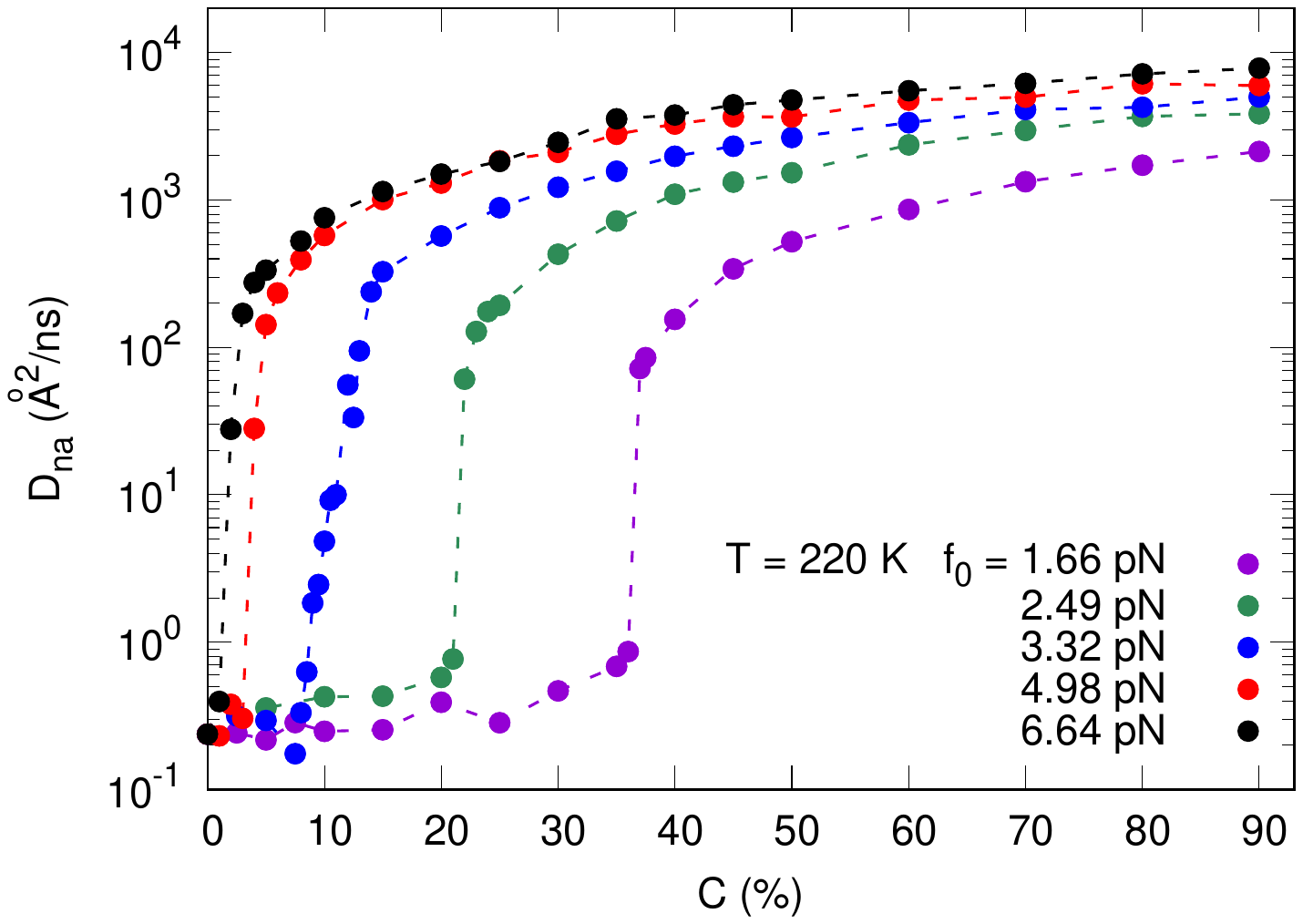}
\caption{(color online) Diffusion coefficient D of non-active molecules as a function of the concentration $C$ of active molecules for various activation  forces. We do not see here the bump that we observed in figure \ref{fig01a} for active molecules.%The dashed line is a guide to the eye that corresponds to $C = 24\%$. %Below the fluidization transition, the diffusion coefficient strongly depends on temperature, whereas above the transition, the data nearly collapse onto a single curve.
}
\label{fig01b}
\end{figure}

We found that facilitation increases with the activation force and the presence of an optimum. 
These results may permit to study the effect of facilitation on the transition. Considering the active molecules in Figure \ref{fig01a} we observe a change in the shape of the curve beginning around the force optimum. We do not see such features for non active molecules (Figure \ref{fig01b}), only a slight change in the slope of the transition. 
Another possible effect is the displacement of the transition critical point on the optimum, the transition being facilitated.
We investigate that question in Figure \ref{fig08} that shows the evolution of the critical concentration with the activation force. 
The critical concentration decreases when we increase the activation force, as expected for an increase of the facilitation.
We also observe a change in the slope of this decrease around the optimum force.}
The evolution of the critical concentration follows two different power laws. For small forces the critical concentration evolves as $C \approx C_0 . (f/f_{0})^{-1.2}$ while for larger forces  $C \approx C_0 . (f/f_{0})^{-2}$.
For small forces the activation modifies only slightly the medium and we expect to be in the linear response theory regime.
Then for larger forces we deviate from this regime as the activation begins to modify permanently the medium.
If they do not show an optimum on the critical concentration, 
the results displayed in Figure \ref{fig08} are however in agreement with our previous interpretation of the origin of an optimum for cooperative motions.
In the linear response regime the facilitation mechanism increases with the force then it decreases when the activation modifies permanently the medium as it does for an increase in temperature.

\begin{figure}%[H]
\centering
\includegraphics[height = 7 cm]{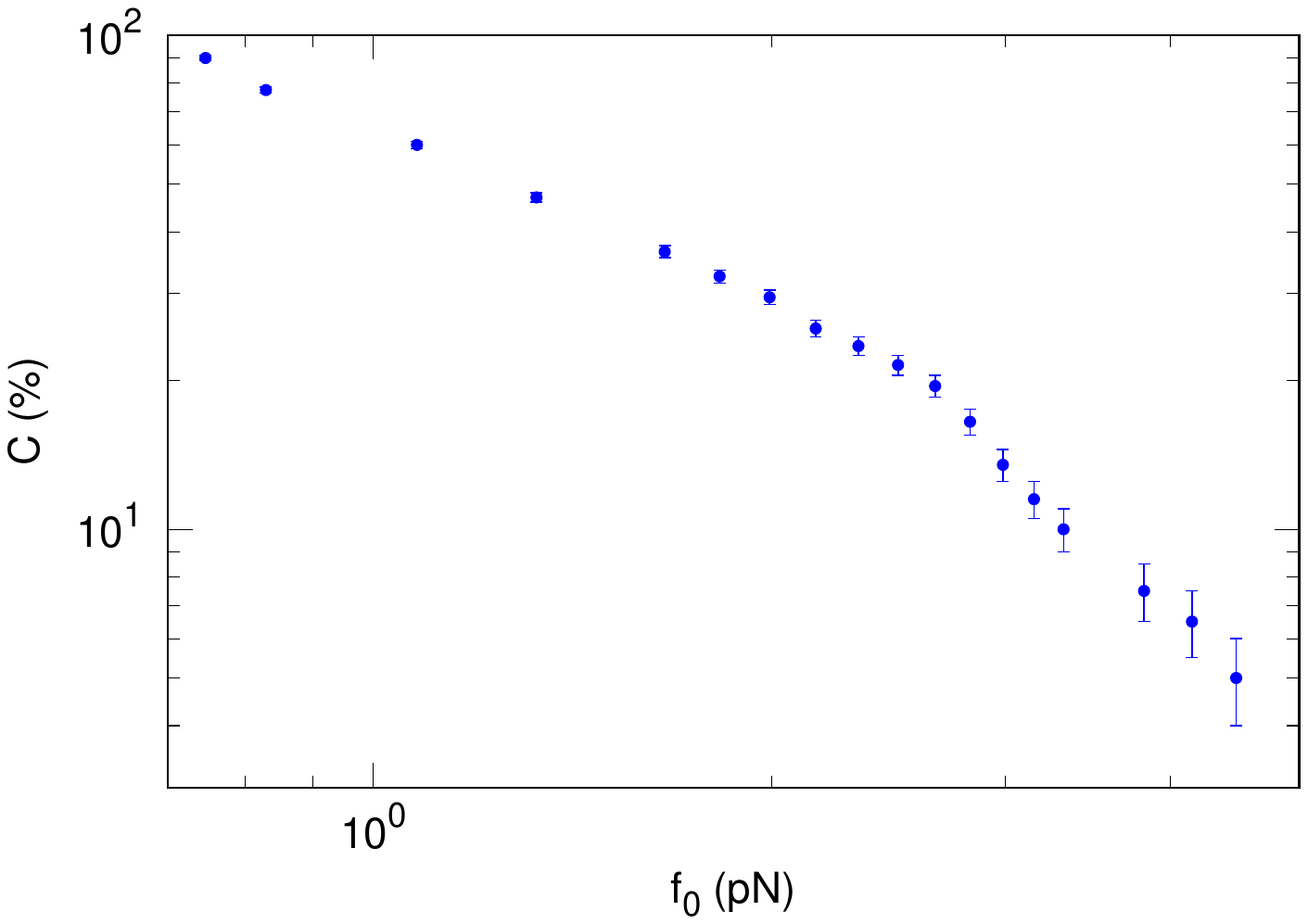}
	
\caption{(color online) Critical active force $f^c$ for the fluidization transition as a function of the concentration of active molecules $C$ at temperature $T = 220 K$. The  dashed line indicates that $f^c \propto C^{-0.5}$ for concentrations ($C \le 30\%$), while the red dashed line indicates that $f^c \propto C^{-0.83}$ for (higher) concentrations.
The change in slope appears around the force $f^c=3.32 pN$ indicating a change in physical regime above this force, that was also observed with cooperativity in previous figures.}
\label{fig08}
\end{figure}

\subsection{A comparison between the effects of force and temperature on the transition}

When we increase the activation force, the facilitation and cooperative motions increase as discussed above.
However the increase of the activation force has a more direct effect on the system as it helps the active molecules overcome the cage energy barrier.
In that view it can be seen as somehow equivalent to an effective temperature increase, for active molecules at least.
However an increase in temperature is expected to decrease the dynamic heterogeneities, in opposition with our results.
%while the force increase has the opposite effect, as it increases facilitation and cooperative motions. 
Also, when the temperature increases, the diffusion below the transition will increase. Therefore above a certain temperature there will be no more transition, also in opposition with what we observe.
%This is in opposition with the force, as increasing the force doesn't decrease the transition.
To better understand the origin of these differences we compare in this section the effects of an increase of the activation force with an increase in temperature.
We will begin by a study of the effect of the activation force on the averaged intermolecular potential energy barrier caused by the cage effect.
Notice that we choose in this work, forces that are not large enough to alone overcome the energy barrier.
Therefore, molecules can only overcome the barrier due to thermal energy fluctuations or the simultaneous presence of several active molecules in the vicinity (that is cooperative motions).\\

\begin{figure}%[H]
\centering
\includegraphics[height = 6 cm]{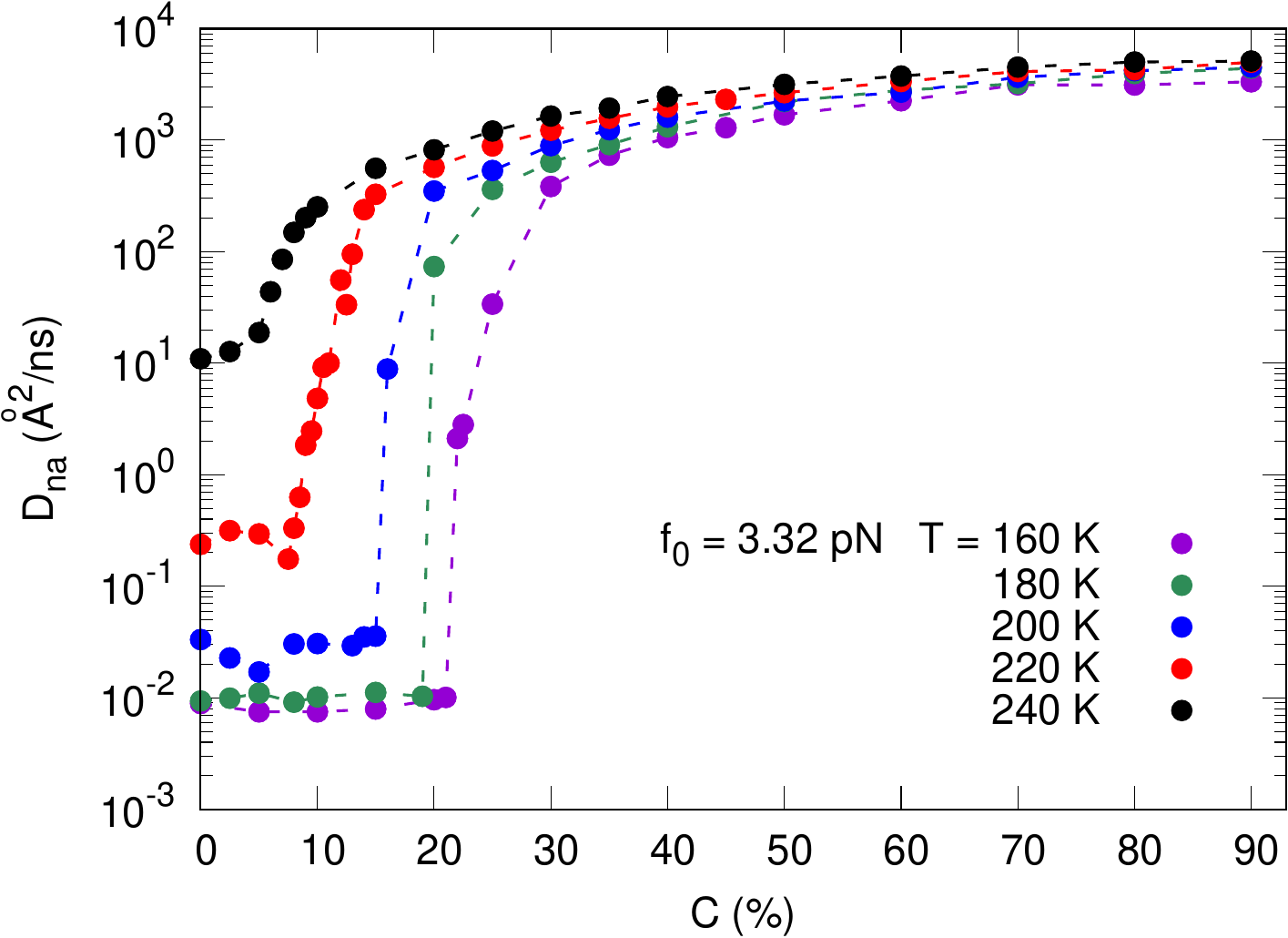}
\caption{(color online) Diffusion coefficient D of non-active molecules as a function of the concentration $C$ of active molecules for different temperatures. %The dashed line is a guide to the eye that corresponds to $C = 24\%$. %Below the fluidization transition, the diffusion coefficient strongly depends on temperature, whereas above the transition, the data nearly collapse onto a single curve.
}
\label{fig02b}
\end{figure}

\begin{figure}%[H]
\centering
\includegraphics[height = 6 cm]{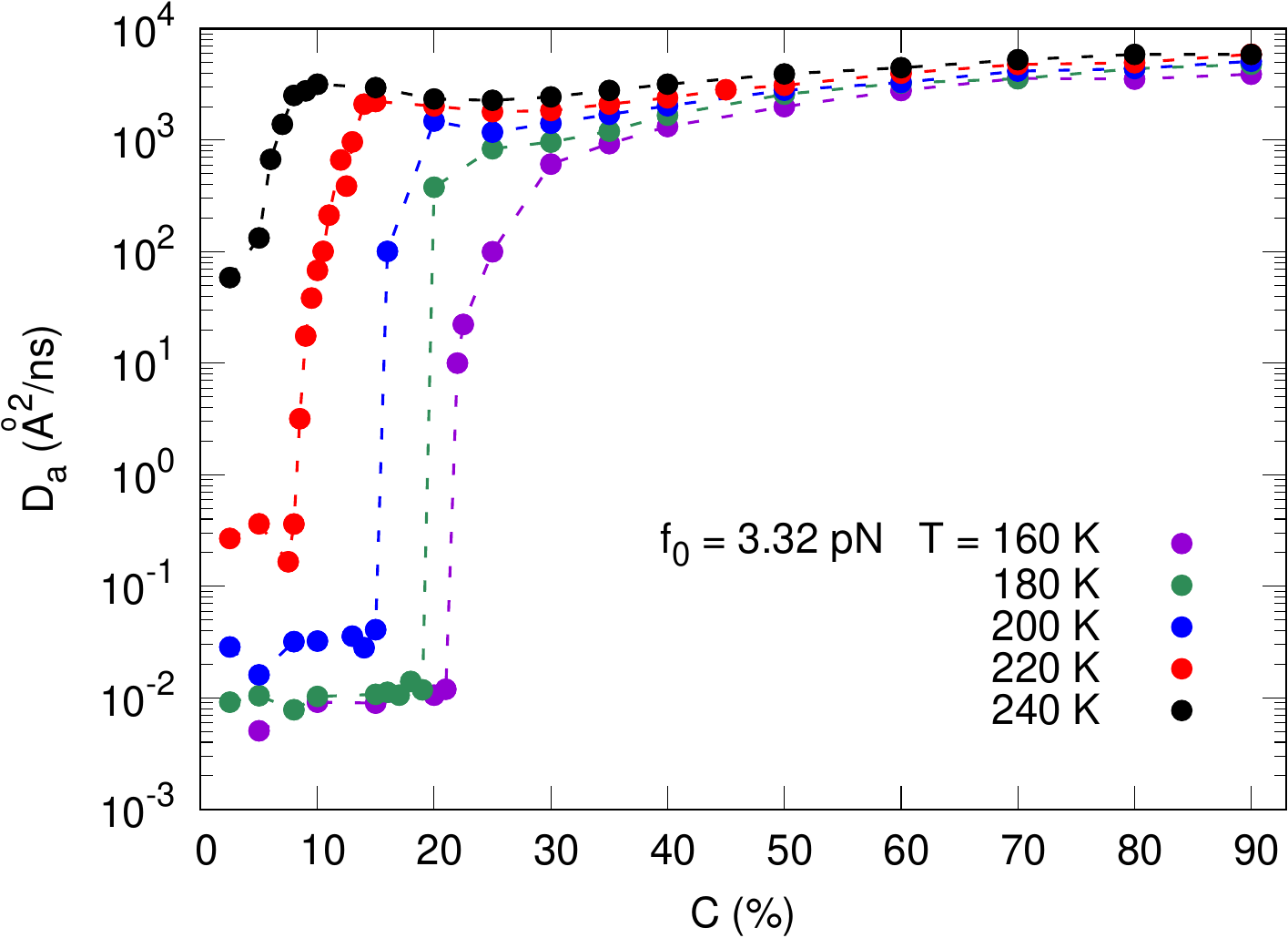}
\caption{(color online) Diffusion coefficient D of active molecules as a function of the concentration $C$ of active molecules for different temperatures. The active and non active plots are similar, but there is a small bump for the active molecules around the transition at low temperatures.%The dashed line is a guide to the eye that corresponds to $C = 24\%$. %Below the fluidization transition, the diffusion coefficient strongly depends on temperature, whereas above the transition, the data nearly collapse onto a single curve.
}
\label{fig02a}
\end{figure}

\begin{figure}%[H]
\centering
\includegraphics[height = 7 cm]{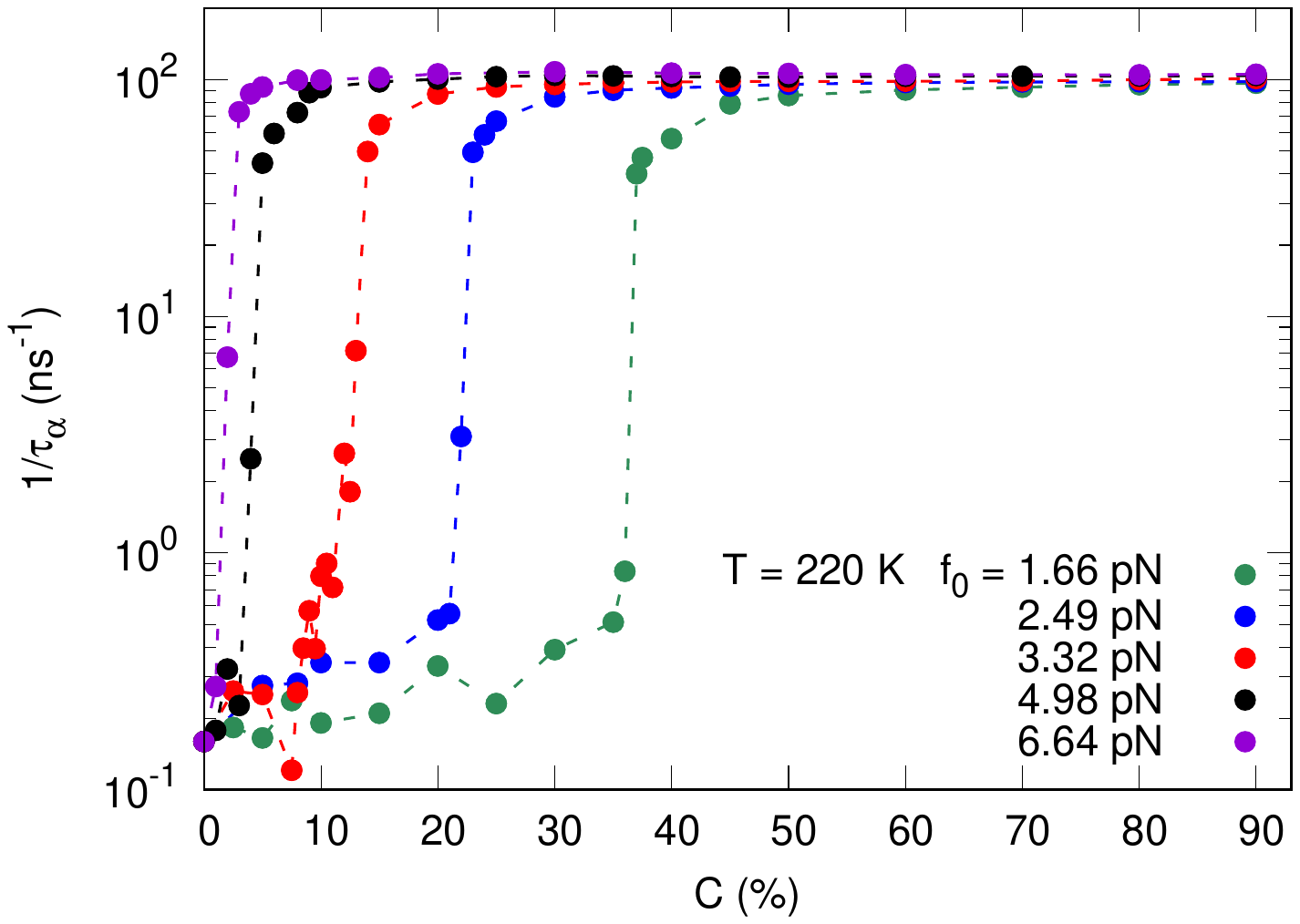}
\caption{(color online) Inverse of the alpha relaxation time $1/\tau_{\alpha}$ of non-active molecules as a function of the concentration $C$ of active molecules for different temperatures. %The dashed line is a guide to the eye that corresponds to $C = 24\%$. %Below the fluidization transition, the diffusion coefficient strongly depends on temperature, whereas above the transition, the data nearly collapse onto a single curve.
}
\label{fig03b}
\end{figure}

 Figures \ref{fig01a}, \ref{fig01b}, \ref{fig02b}, \ref{fig02a}, and  \ref{fig03b} show the evolution of the transition when the activation force or the temperature increases.
 We observe similarities between the two evolutions.
 In both cases, the temperature as the force shift the transition to smaller critical concentrations, due to the decrease of the energy barrier to be overcomed. 
  However we observe that the diffusion above the transition, depends on the force in Figures \ref{fig01a} and \ref{fig01b} while it doesn't depend on the temperature in Figures \ref{fig02b} and \ref{fig02a}.
 We interpret the force dependence of the diffusion above the transition to be due to an increase in the facilitation mechanism leading to an increase in diffusion.
 The relaxation time behavior for various forces in Figure \ref{fig03b} confirms that interpretation as we observe that it follows a similar trend than the temperature behavior. %Therefore the difference comes from the breaking of Stokes-Einstein law between diffusion and the relaxation time, an effect usually attributed to cooperative motions. 
 Actually, above the transition, the relaxation time doesn't depend on the force. The dependence observed for the diffusion therefore comes from the divergence between $1/\tau_\alpha$ and $D$, that is the breaking of Stokes-Einstein law that is usually attributed to the rise of cooperative motions.\\
 
 Below the transition, at small concentrations, we observe the opposite behavior, as the diffusion depends on the temperature but not on the force.
The diffusion dependence on the temperature comes from the Boltzmann variation of the concentration of excitations.
 This effect doesn't appear for the force as the force is only acting on active molecules.
 It therefore indicates that the force induced motion is only transmitted to non-actives above the transition.
 %We then expect differences arising from the effect of cooperative motions.
%To test this interpretation we display the direct effect on active molecules in Figures \ref{fig01a} and \ref{fig02a}.

 \vskip 0.5cm
%From the two plots we may roughly define an effective temperature, as the temperature shift equivalent to the force increase.
%However while the force increases facilitation and cooperative motions, the temperature has the opposite effect.
%The difference between the two figures comes then mostly from cooperative motions.
%\\

\section{Conclusion}

 In this work we studied the effect of varying the activation force on the fluidization transition.
 Increasing the activation force in active matter leads to an increase of loose structures and as a result to an increase in facilitation mechanisms and associated cooperative motions. Our aim was to be able to separate the facilitation effect at the origin of the fluidization transition from more direct effects, as for example the probability increase for active molecules to overcome the cage energy barrier, an effect that could be seen as a local effective temperature increase.
 
 We actually found an increase of cooperative motions when the activation force increases.
 We found peaks of dynamic heterogeneities (cooperative motions) on the transition.
 If peaks of cooperative motions are expected for a phase transition\cite{book1,book2,book3}, 
 the peaks of dynamical heterogeneities (DH) on the transition suggest that the DH are the cooperative motions physically significant for the transition.
 
 Studying the DH evolution with the activation force, we found an optimum force  for cooperative motions and facilitation.
 For that value we didn't find however any significant  modification on the transition, but a bump developed on active particles diffusion, showing an increase of diffusion.
 
 As the force increases the probability to overcome the cage energy barrier as does an increase of temperature, we then compared the force effect on the transition with the effect of temperature. We found in both cases a shift of the transition to smaller critical concentrations. 
 However a temperature increase is associated to larger energies given to the medium, and to a decrease of cooperative motions while the entropy increases.

We finally  interpret the transition as originating from the increase of probability to overcome the energy barrier when the force or the concentration increases,
 while the cooperative mechanism permits active molecules to transmit the diffusive motions to the whole medium.

%L'accroissement de la coopérativité et la décroissance du mur du potentiel expliquent la fluidisation. La transition de phase est expliquée par la transmission de la coopérativité, de la dynamique/chute du mur, de la concentration d'excitations, à l'ensemble des molécules.

%Nous espérons que cela pourra aider à la compréhension de la transition liquide-verre. Cependant dans la transition liquide verre, contrairement à la matière active, la coopérativité décroît lorsque la densité d'excitations s'accroît.

%\vskip 0.5cm

%{\bf \Large  Conflict of interest}

%There are no conflict of interest to declare.

%\vskip 0.5cm

%{\bf \Large  Data availability}

%The data that support the findings of that study are available from the corresponding author upon reasonable request.

%\vskip 0.5cm

%\newpage

\end{document}